\documentclass{article}

\usepackage{microtype}
\usepackage{graphicx}
\usepackage{subfigure}
\usepackage{booktabs} 
\usepackage{amsmath}

\usepackage{mathtools}          
\usepackage{amssymb}

\usepackage{hyperref}

\usepackage[accepted]{icml2019}

\icmltitlerunning{None}

\begin{document}

\twocolumn[
\icmltitle{Dynamically Expanded CNN Array for Video Coding}



\icmlsetsymbol{equal}{*}

\begin{icmlauthorlist}
\icmlauthor{Everett Fall}{ntu}
\icmlauthor{Kai-Wei Chang}{ntu}
\icmlauthor{Liang-Gee Chen}{ntu}
\end{icmlauthorlist}

\icmlaffiliation{ntu}{National Taiwan University}

\icmlcorrespondingauthor{Everett Fall}{everett.fall@gmail.com}

\icmlkeywords{Dynamically expanding, CNN array, CNN ensemble, CNN video refinement, video coding}

\vskip 0.3in
]




\begin{abstract}
\label{sec:algorithm}
Video coding is a critical step in all popular methods of streaming video. Marked progress has been made in video quality, compression, and computational efficiency.
Recently, there has been an interest in finding ways to apply techniques form the fast-progressing field of Machine Learning to further improve video coding.

We present a method that uses convolutional neural networks to help refine the output of various standard coding methods.
The novelty of our approach is to train multiple different sets of network parameters, with each set corresponding to a specific, short segment of video.
The array of network parameter sets expands dynamically to match a video of any length. 
We show that our method can improve the quality and compression efficiency of standard video codecs.

\end{abstract}


\section{Introduction}
\label{sec:introduction}

In recent years there have been many advances in video coding standards, and there are several prominent video codecs used commercially such as H.264/AVC \cite{DBLP:journals/tcsv/SchwarzMW07}, H.265/HECV \cite{DBLP:journals/tcsv/SullivanOHW12} and VP9.
However, most of these traditional coding algorithms are block-based and therefore suffer from block artifacts.
They are also largely hand-designed, making joint optimization difficult.

In Machine Learning (ML) the past decade has yielded vast improvements and altogether new techniques for processing images and video such as convolutional neural networks (CNN). 
CNNs offer an intuitive theoretical approach of encoding or \emph{embedding} information as vectors, that can be interpreted as high-level features.
Despite the success of CNNs in many ML tasks, they have seen very limited success as a direct replacement for traditional commercial video codecs.
Partially because training ML models is computationally expensive and partially because existing methods such as .H265 or VP9 are already highly optimized, making the improving on benchmarks a challenging barrier to entry. 

In this work we take a different approach. Instead of directly replacing traditional codecs by using CNNs to learn an end-to-end model, we apply ML as a post processing step to refine the output.
The novelty of our approach is to train a dynamically expanding array of \emph{many small CNNs}, allowing each network to specialize in refining a relatively short segment of video.
This refiner module switches between different networks as the video is decoded.

We evaluate our method on several commonly used commercial codecs and show that the refiner network makes a substantial improvement to quality with only a minimal increase in code size.


\section{Related Work}
\label{sec:relatedwork}

Several deep learning-integrated video compression methods have been proposed to improve traditional coding.
One approach is replacing and enhancing different modules in traditional coding, especially the state-of-the-art HEVC codec. For example,
improving motion compensation and inter-prediction \cite{DBLP:conf/iscas/HuoLWL18, DBLP:journals/tcsv/YanLLLLW19, DBLP:conf/icip/ZhaoWZWM018},
intra-prediction \cite{DBLP:conf/vcip/SongLLW17, DBLP:journals/corr/abs-1808-05734, zxcvasdf}, and
replacing in-loop filter \cite{DBLP:conf/ivmsp/ParkK16, DBLP:conf/icip/KangKL17, DBLP:journals/tip/ZhangSJZXD18, 8630681}.

Another approach is to apply ML as a post processing step to improve video quality. For example, \cite{DBLP:conf/icip/LiSXZ17} proposed a dynamic metadata post-precessing scheme based on a CNN, \cite{DBLP:conf/icmcs/YangXW17} and \cite{DBLP:conf/dcc/WangCC17} proposed Decoder-side Scalable  Convolutional Neural Network (DS-CNN) and Deep CNN-based Auto Decoder (DACD) respectively for video qulity and efficiency enhancement.

Instead of using ML to form a hybrid video coding framework, some works propose an end-to-end framework for video compression, the performance of which can be on par with the commercial codecs. \cite{DBLP:conf/eccv/WuSK18} developed an end-to-end deep video codec relying on repeated interpolate images in a hierarchical manner. Inheriting conventional video coding structure, \cite{DBLP:journals/corr/abs-1812-00101} employed multiple neural networks to constitute different modules, which can be jointly optimized through a single loss function. \cite{DBLP:journals/corr/abs-1811-06981} proposed a learned end-to-end model for low-latency mode with spatial rate control.


\section{Dynamically Expanded CNN Array}
\label{sec:algorithm}

We denote a video as a sequence of frames: $X={x_1,x_2,...}$ where each $x_t \in \mathbb{R}^{W \times H \times 3}$ is an image with width $W$, height $H$ with 3 color channels for each pixel. 
An video coding scheme provides two algorithms: one to transform the video into code, $E: \mathcal{X}\mapsto\mathbb{R}^N$, and one that converts an encoded video back into a sequence of images, $D: \mathbb{R}^N \mapsto \mathcal{X}$. 
Let $\hat{X}\in\mathcal{X}$ denote a sequence of frames generated by encoding and then decoding a video: $\hat{X}=D(E(X))$.
The goal of video coding is to minimize size of an encoded video and to design $E$ and $D$ to be as computationally efficient as possible.
In the case of lossy compression coding, there is a pixel-wise error, known as the \emph{residual}, associated with each decoded frame $\Delta_t=\hat{x_t}-x_t$, which is also desirable to minimize. 

\begin{figure}[ht]
\vskip 0.2in
\begin{center}
\centerline{\includegraphics[width=\columnwidth]{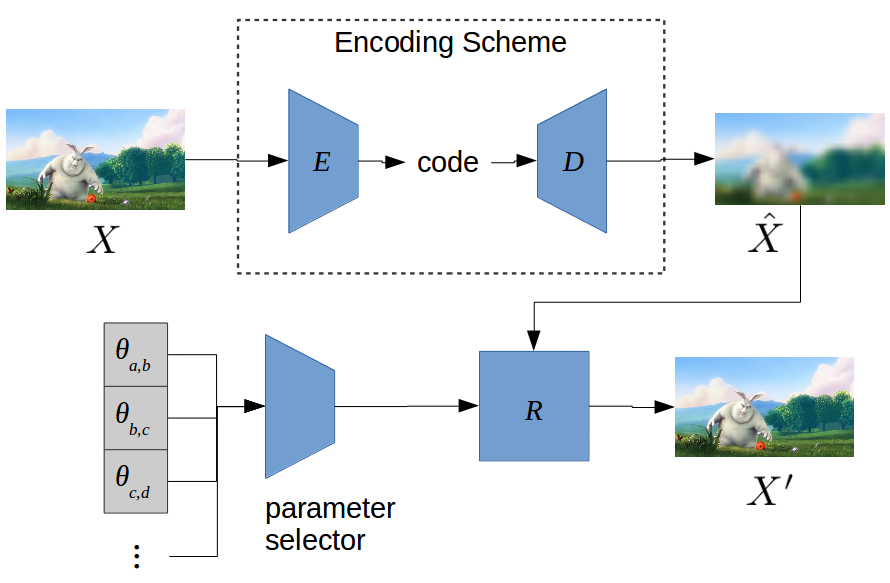}}
    \caption{The video $X$ is encoded and decoded by some coding scheme to produce $\hat{X}$. The correct parameters for the refiner network are selected according to current frame being process. The refiner produces $X'$ from $\hat{X}$.}
\label{flow_diagram}
\end{center}
\vskip -0.2in
\end{figure}

Our method is designed to complement an existing coding scheme, reducing the error in the decoded video by adding a refining function $x'_t=R(\hat{x}_t): \mathbb{R}^{W\times H\times 3}\mapsto \mathbb{R}^{W\times H\times 3}$ which yields $X'=x'_1,x'_2,...$ when applied to each frame of $X$ as shown in Fig. \ref{flow_diagram}. 
The existing coding scheme could be a standard commercial codec or a custom end-to-end CNN.

Let $s_{i,j}={x_i,x_{i+1},...,x_j}$ denote a contiguous segment of $X$. 
The novelty of our method is to use a CNN to implement $R$ with parameters $\theta(t)$, which vary with time.
Specifically, we partition $X$ in to many small segments of duration $\rho$, $s_{1,\rho},s_{\rho+1,\rho \times 2},...$, and for each segment a corresponding set of network parameters is learned for $R$ which we denote as $\theta_{i,j}$.
Intuitively, this can be thought of as an array of CNNs that expands dynamically as needed to refine a video of any length.

Learning $\theta_{i,j}$ is accomplished through standard stochastic gradient decent.
A training example consists of $(x=\hat{x}_t,y=x_t)$ where $x$ is sampled from $X$ in the range $i\leq t\leq j$.
Intuitively the network learns to predict $x_t$ from $\hat{x}_t$.
Alternatively, the refine function can learn to predict the residual (denoted $\Delta'_t=R_{\Delta}(\hat{x}_t)$) in which case refining involves removing the predicted residual from the signal.
In this case the training example consists of $(x=\hat{x}_t,y=\Delta_t)$ and the output is obtained by $x'_t=\hat{x}_t-R_{\Delta}(\hat{x}_t)$.

Another commonly desired characteristic of an encoding scheme is for the code to have a localized temporal correspondence to the video.
This allows a small segment, known as a \emph{random access segment}, of the video to be decoded without requiring the entire code which is useful for data transmission applications such as video streaming.
Since each segment $s_{i,j}$ is relatively short, our method can also support random access by transmitting $\theta_{i,j}$ in advance of the corresponding random access segment of the code.

\section{Experimental Evaluation}
\label{sec:evaluation}

\begin{figure}[ht]
\vskip 0.2in
\begin{center}
\centerline{\includegraphics[width=\columnwidth]{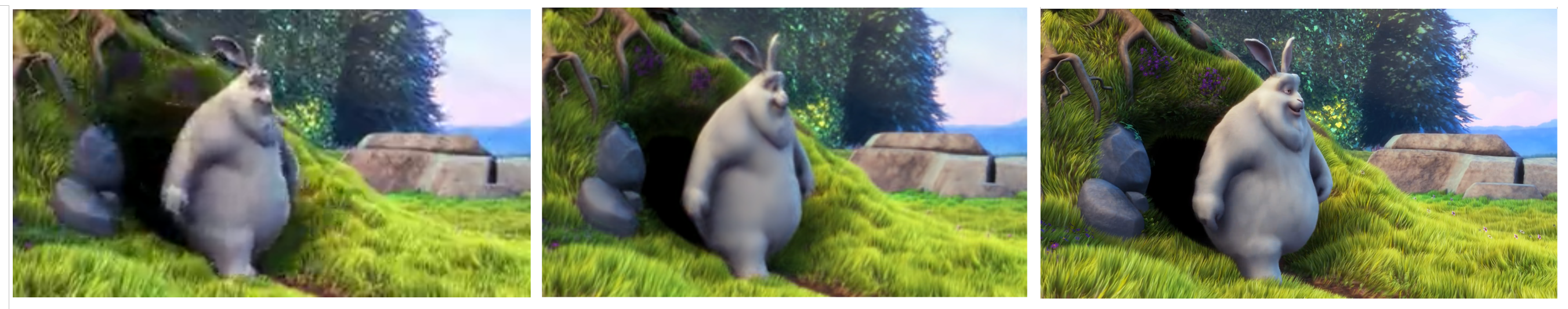}}
    \caption{Left: The refiner input (output of .H264 codec using CRF 36). Middle: Output of the refiner. Right: Ground truth.}
\label{before_after}
\end{center}
\vskip -0.2in
\end{figure}

In this section we present the results of initial experiments as a proof-of-concept for our method. We implemented our proposed method and applied it to the benchmark dataset "Big Buck Bunny".
We used the .H264 codec with a high CRF value (low quality and high compression ratio) to create the input to the refiner network.
The refiner contains 4 convolutional layers with 5x5 filters followed by 3 convolutional layers with 3x3 filters and applies to segments of size $\rho=50$ frames.
The network is given approximately 500 training steps which corresponds to 10 epochs of the (very small) dataset for each 50 frame segment.
The before and after result of a typical frame is shown in Fig. \ref{before_after}. The quality of the refined image was MS-SSIM: 0.9802
PSNR: 36.96 (original .H264 CRF-36 was MS-SSIM: 0.9589 and PSNR: 33.82).

\section{Conclusion}
\label{sec:conclusion}

In this work we introduced a novel method for video coding which uses an array CNNs to refine each frame.
We describe the algorithms used to segment the video and train a CNN for each segment and automatically switch between networks during the decoding process.
We implemented our proposed algorithm and conducted several experiments to evaluate it's performance with standard benchmark coding schemes.
Our method was able to provide substantial improvement to the quality and reduce the compressed video size.

\bibliographystyle{icml2019}
\bibliography{bibs/nn}

\end{document}